\documentclass[%
 aip,
 amsmath,amssymb,
 reprint,%
]{revtex4-1}

\usepackage{graphicx}
\usepackage{dcolumn}
\usepackage{bm}

\usepackage[utf8]{inputenc}
\usepackage[T1]{fontenc}
\usepackage{mathptmx}
\usepackage{etoolbox}

\usepackage{ulem}
\usepackage{color}

\makeatletter
\def\@email#1#2{%
 \endgroup
 \patchcmd{\titleblock@produce}
  {\frontmatter@RRAPformat}
  {\frontmatter@RRAPformat{\produce@RRAP{*#1\href{mailto:#2}{#2}}}\frontmatter@RRAPformat}
  {}{}
}%
\makeatother
\begin{document}

\preprint{AIP/123-QED}

\title{Autler-Townes splitting in the trap-loss fluorescence spectroscopy due to single-step direct Rydberg excitation of cesium cold atomic ensemble}
\author{Xin Wang}

\author{Xiaokai Hou}
\author{Feifei Lu}
\author{Rui Chang}
\author{Lili Hao}
\author{Wenjing Su}
\author{Jiandong Bai}%
 \altaffiliation{Department of Physics,North University of China,Taiyuan 030051,China}
\affiliation{ 
State Key Laboratory of Quantum Optics and Quantum Optics Decices, Shanxi University, Institute of Opto-Electronics,Taiyuan 030006, China
}%

\author{Jun He}%
\author{Junmin Wang}
 \homepage{Corresponding author: wwjjmm@sxu.edu.cn; ORCID : 0000-0001-8055-000X}
\affiliation{ 
State Key Laboratory of Quantum Optics and Quantum Optics Decices, Shanxi University, Institute of Opto-Electronics,Taiyuan 030006, China
}%
\affiliation{%
Collaborative Innovation Center of Extreme Optics, Shanxi University, Taiyuan 030006,China
}%

\date{\today}

\begin{abstract}
We experimentally investigate trap-loss spectra of the cesium $6S_{1/2}(F=4)\rightarrow 71P_{3/2}$ Rydberg state by combining the cesium atomic magneto-optical trap with the narrow-linewidth, continuously-tunable 318.6 nm ultraviolet laser. That is, the atoms in the magneto-optical trap are excited to the Rydberg state due to the ultraviolet laser single-step Rydberg excitation, which leads to the reduction of atomic fluorescence. Based on the trap-loss spectroscopy technology, the Autler-Townes splitting due to strong cooling laser is observed, and the parameter dependence of the AT splitting interval of trap-loss spectroscopy is investigated. Furthermore, the effective temperature of cold atoms is measured by means of simplified time-of-flight fluorescence imaging. In addition, closed-loop positive feedback power stabilization of 318.6 nm ultraviolet laser is carried out. This lays the foundation for further experimental research related to Rydberg atoms using ultraviolet lasers, which is of great significance for the development of quantum computing and quantum information fields. 
\end{abstract}

\maketitle

\section{\label{sec:level1}Introduction}

The stronger long-range dipole-dipole interactions between highly excited Rydberg atoms resulting in Rydberg blockade. It is very promising for applications in multi-body physics$^{1}$, quantum computing$^{2}$, quantum information$^{3}$, nonlinear optics$^{4-5}$, and imaging$^{6-7}$. For the detection of Rydberg atoms, optical detection$^{8}$ and field ionization detection$^{9}$ are generally adopted. For the applications of quantum information, non-destructive detection is necessary. Therefore, the method of all-optical detection using the step-type electromagnetically induced transparency (EIT) spectra of Rydberg atoms$^{10-11}$ is widely used. Experimentally, highly excited Rydberg atoms are usually prepared by cascaded two-photon or three-photon excitation. The single-photon excitation can avoid atomic decoherence from population in the intermediate state, the photon scattering, the AC-Stark shift during multi-photon excitation. Therefore, the single-photon excitation scheme has obvious advantages for the preparation of Rydberg atoms for quantum computing and quantum information.

 The use of single-step excitation to prepare Rydberg state atoms has a low probability of direct excitation, and the transition wavelength is generally in the violet or ultraviolet (UV) band, which is not easy to achieve, so there are fewer experiments using single-photon excitation to prepare Rydberg state atoms. In 2004, Tong et al. obtained a 297 nm UV pulsed laser by doubling the frequency of a 594 nm dye laser and achieved single-photon Rydberg excitation of $^{85}$Rb atoms $5S_{1/2}\rightarrow nP_{3/2}$(n = 30-80) in rubidium cold atomic magneto-optical trap$^{12}$. In 2009, Becker's group used a similar device to obtain a 297 nm UV continuum laser to achieve single-photon Rydberg excitation of the $^{85}$Rb atom $5S_{1/2}\rightarrow 63P_{3/2}$ in a room atomic vapor cell$^{13}$. In 2019, Whitlock's group used a 572 nm dye laser to doubling the frequency to produce a 286 nm laser for the experimental study of the $^{39}$K cold-atom Rydberg-dressed-ground-state Ramsey interferometer$^{14}$.

In recent years, with the development of nonlinear optical frequency conversion technology and quasi-phase matching technology, as well as the maturity of crystal materials and crystal coating technology, the implementation of continuously tunable UV laser has been gradually developed. In 2014, the Biedermann's group used the sum frequency of 1071 nm and 1574 nm to generate a 638 nm laser, after which a 300 mW 319 nm continuous UV laser was obtained by frequency doubling and used for single-photon Rydberg excitation of cesium atoms. The Rydberg blocking effect was observed in two single-atom optical dipole traps at a distance of 6.6 $\mu$m $^{15}$. In 2019, Li Xiaolin's group obtained a 297 nm ultraviolet laser of about 200 mW by quadrupling the frequency of 1188 nm infrared laser, which was used for the experimental study of single-step Rydberg excitation in the rubidium hot atomic vapor cell$^{16-17}$. In 2020, our experimental group explored the DC electric field sensing of cesium cold atomic systems using a 319 nm UV laser$^{18}$.

In the paper, single-step Rydberg excitation in cesium cold atomic ensemble is achieved by using a narrow-linewidth single-frequency 318.6 nm ultraviolet laser and all-optical detection scheme. In the experiment, 637.2 nm red light is generated by the sum frequency of 1560.5 nm and 1076.9 nm, and then 318.6 nm UV laser of ~2W is generated by the double frequency of red light. Then a cesium magneto-optical trap is constructed to measure the effective temperature, size and atomic density of cold atoms by means of simplified time-of-flight fluorescence imaging. Finally, the trap-loss spectra of $6S_{1/2}(F=4)\rightarrow 71P_{3/2}$ Rydberg state is studied by 318.6 nm UV laser combined with cesium magneto-optical trap. Based on the trap-loss spectroscopy technology, the Autler-Townes (AT) splitting due to strong cooling laser is observed, and the parameter dependence of the AT splitting interval with trap-loss spectroscopy is investigated. The single-photon Rydberg excitation in this paper has positive implications for the further development of quantum optics and quantum information processing using cold atomic samples.

\section{The vapor-cell magneto-optical trap of cesium atoms}

The specific Cs atomic MOT schematic is shown in Fig. 1(a), where the MOT is loaded in a vacuum glass cell with size of 30 mm $\times$ 30 mm$\times$ 120 mm, with wall thickness of 5 mm and vacuum degree maintained at $\sim$ 10$^{-10}$ torr. In figure, the quadrupole magnetic field gradient required for the MOT is provided by a pair of inverted helmholtz coils, as shown in the gray coil, which are fixed in front (-z) and behind (z) of the glass cell, producing a magnetic field gradient of 32.0 Gauss/cm when the current is 1.6 A.

The 852-nm cooling laser is provided by a grating feedback external cavity diode laser (ECDL) of Wavicle with output power of $\sim$ 90 mW, and then through the laser amplifier of Toptia can produce 852 nm laser of $\sim$ 200 mW, and beam diameter of $\sim$ 10 mm. The cooling laser is detuned by $\Delta$$_{12}$=-12.4 MHz from the Cs $6S_{1/2}(F=4)\rightarrow 6P_{3/2}(F'=5)$ transition. The 852-nm repumping laser is provided by a distributed-Bragg-reflector (DBR) diode Laser with output power of $\sim$ 80 mW. The repumping laser resonance at the Cs  $6S_{1/2}(F=3)\rightarrow 6P_{3/2}(F'=4)$transition. and beam diameter of $\sim$ 11.5 mm. The Angle of cooling laser and repumping laser in the XY plane is 30$^{\circ}$. As shown in Fig. 1(a), the six cooling laser and repumping laser overlap in the cesium atomic vacuum glass cell, and the intersection point coincides with the zero point of the magnetic field of anti-helmholtz coil. The effective temperature, cold atomic size and atomic number density of cesium magneto-optical trap are measured by simplified time-of-flight fluorescence imaging. The relevant energy levels are shown in Fig.1(b).

For the temperature measurement of cold atomic samples, we use a simplified time-of-flight fluorescence imaging method, specifically, the cooling laser for laser cooling and trapping atom instead of additional probe laser of standard time-of-flight fluorescence imaging. The inset shown in Fig. 2 can be obtained as a grayscale of the fluorescence image with time intervals of 2.1, 4.0, 6.0, 8.0, and 10.0 ms, respectively. The atomic Gaussian radii are obtained by processing different atomic fluorescence images as a function of diffusion time.

\begin{figure}
\includegraphics[width=8cm]{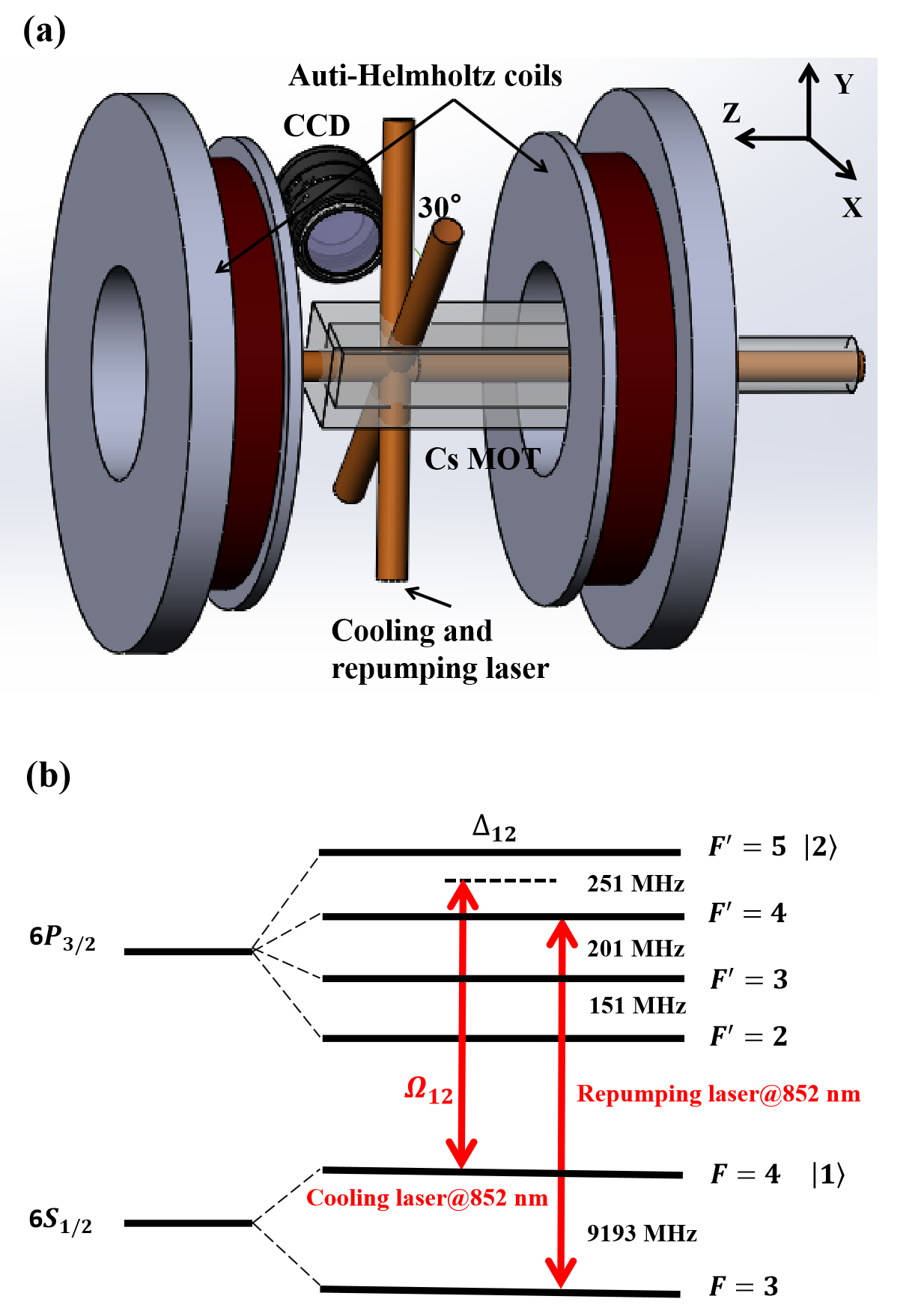}
\caption{\label{fig:epsart} Relevant hyperfine levels for Cs atomic MOT. (a) Schematic diagram of cesium atomic magneto-optical trap; (b) Energy level diagram, the cooling beams have Rabi frequency $\Omega_{12}$. $\Delta_{12}$ is the detuning of cooling laser, the 852-nm cooling laser is detuned by $\Delta_{12}$ from the Cs  $6S_{1/2}(F=4)\rightarrow 6P_{3/2}(F'=5)$ transition, the 852-nm repumping laser resonance at the Cs  $6S_{1/2}(F=3)\rightarrow 6P_{3/2}(F'=4)$ transition. }
\end{figure}

\begin{figure}
\includegraphics[width=8cm]{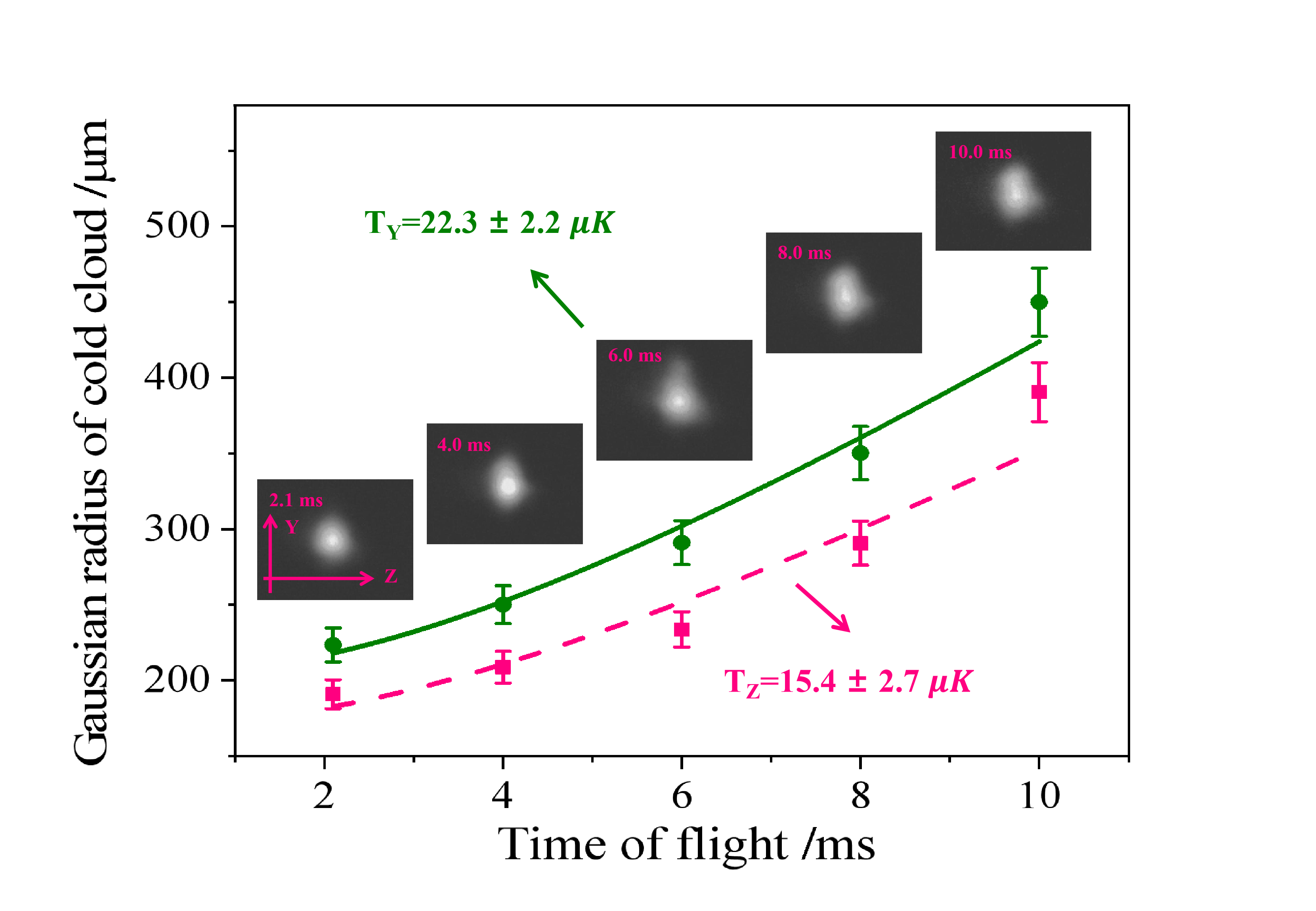}
\caption{\label{fig:epsart} Fitting of effective temperature data for cold atoms, green is the atomic Gaussian radius measured in the Z-direction with time of flight, and the fitted effective temperature is 15.4 $\pm$ 2.7 $\mu$K, pink is the atomic Gaussian radius measured in the Y-direction with time of flight, and the fitted effective temperature is 22.3 $\pm$ 2.2 $\mu$K.}
\end{figure}

Using the relationship between the rate of diffusive expansion of the atomic cloud and the temperature:$\sigma^2_t$= $\sigma^2_0$+ $\frac{k_BT}{m}$$t^2$ , the initial temperature of the atom and the initial size of the atomic cloud can be obtained after the fit, where m is the atomic mass (for $^{133}$ Cs atoms m = 2.2$\times$ 10$^{-25}$ Kg), $k_B$ = 1.38$\times$ 10$^{-23}$ J/K is the Boltzmann constant, T is the effective temperature of the cold atom, $\sigma_t$ is the Gaussian radius of the atomic cloud at time t, and $\sigma_0$ is the initial radius of the cold atomic cloud.

As shown in Fig.2, along the Z-direction, the atomic diameter is 365.4 $\pm$ 3.2 $\mu$m and the effective temperature is 15.4 $\pm$ 2.7 $\mu$K, while along the Y-direction the atomic diameter is 436.3 $\pm$ 4.5 $\mu$m and the effective temperature is 22.3 $\pm$ 2.2 $\mu$K. 
Furthermore,the average density of cold atoms can be estimated to be $\sim$2.2$\times$10$^{10}$ cm$^{-3}$. Note, The experimentally measured temperature is consistent with common sense, and the Doppler cooling limit temperature of the cesium atom is 125 $\mu$K. Even without polarization gradient cooling in MOT, there is a sub-Doppler cooling mechanism that can make the temperature of the atoms in MOT lower than the Doppler cooling limit temperature.

\begin{figure*}
\includegraphics[width=16cm]{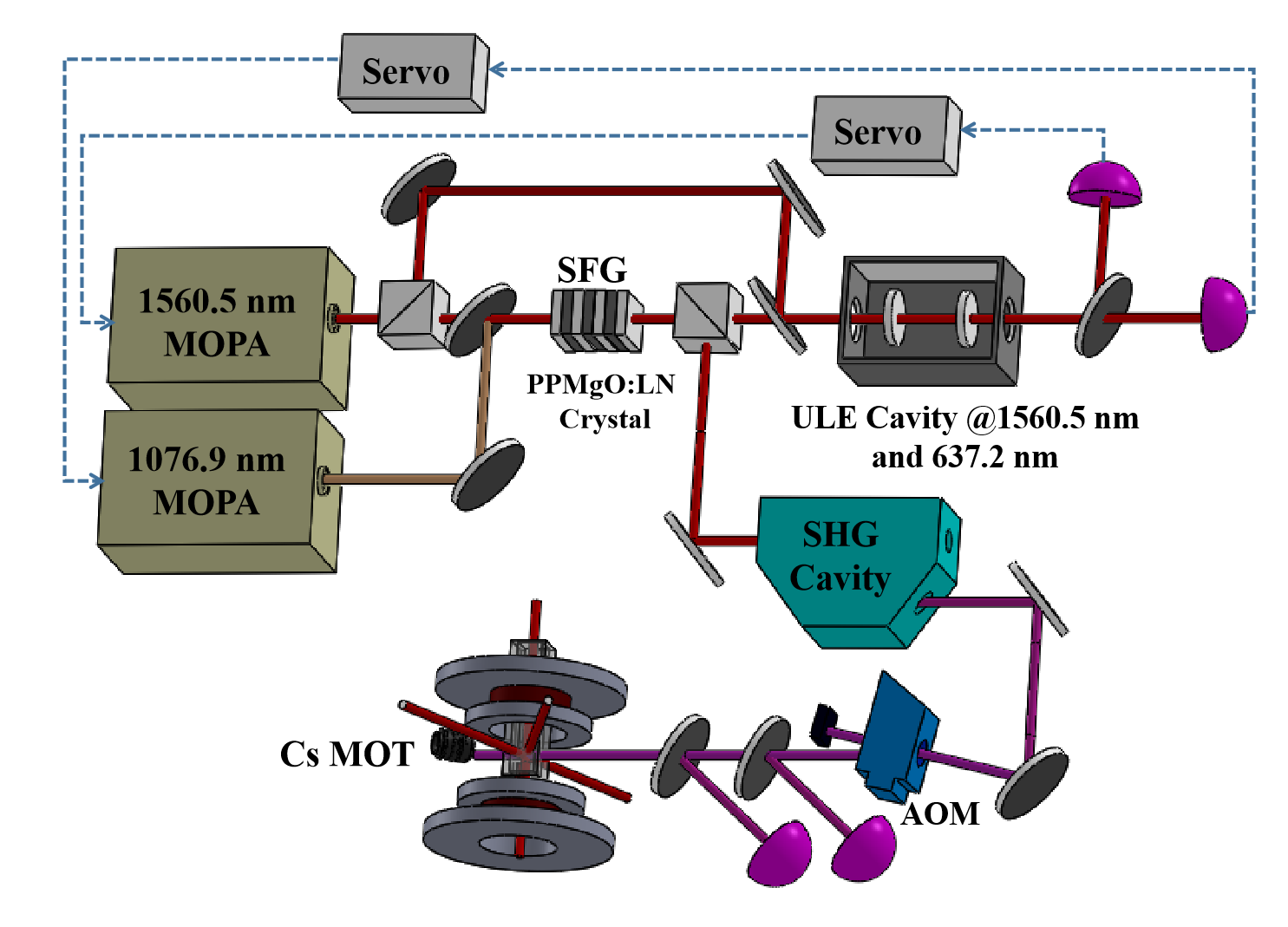}
\caption{\label{fig:wide}Diagram of the experimental setup, where the 1560.5 nm MOPA system consists of a distributed feedback erbium-doped fiber laser (DFB-ErDFL) with a narrow linewidth (160 Hz) of 1560.5 nm and an erbium-doped amplifier (ErDFA), between which a waveguide-type electro-optical phase modulator (EOPM) with an input-output polarization-preserving pigtail is inserted; 1076.9 nm MOPA system consists of a distributed feedback ytterbium-doped fiber laser (DFB-YbDFL) with a narrow linewidth (2 kHz) of 1076.9 nm and an ytterbium-doped amplifier (YbDFA); AOM closed-loop positive feedback power stabilization device; Cesium atomic magneto-optical trap including auti-helmholtz coil, vacuum glass cell.}
\end{figure*}

\section{Ultraviolet laser system and power stability}

The experimental setup is shown in Fig.3, with references to the literature $^{19-21}$ for details. The 1560.5 nm MOPA system consists of a 1560.5 nm distributed feedback erbium-doped fiber laser (DFB-ErDFL) with output power of $\sim$ 200 mW and linewidth of $\sim$ 160 Hz. It is used as a seed laser injected an erbium-doped amplifier (ErDFA) with wavelength of 1540 $\sim$ 1565 nm. The beam is $\sim$ 1.4 mm in diameter and can produce a nominal 1560.5 nm laser of $\sim$ 15W. A waveguide-type electro-optic phase modulator (EOPM) is inserted between the 1560.5 nm laser and the amplifier for phase modulation. The 1076.9 nm MOPA system consists of a 1076.9 nm distributed feedback ytterbium-doped fiber laser (DFB-YbDFL) with output power of $\sim$ 80 mW and linewidth of $\sim$ 2 kHz. It is used as a seed laser injected an ytterbium-doped amplifier (YbDFA) with wavelength of 1060 $\sim$ 1090 nm. The beam is $\sim$ 1.7 mm in diameter and can produce a nominal 1076.9 nm laser of $\sim$ 10 W. As can be seen from the figure, the 1560.5 nm laser and 1076.9 nm laser pass through the periodically polarized PPMgO:LN (PPLN)crystal, and the sum frequency to produce 637.2 nm red light, after which the 637.2 nm laser is injected into the four-mirror ring doubling cavity to produce $\sim$ 2 W of narrow-linewidth, continuously-tunable 318.6 nm UV laser, and the doubling crystal is BBO crystal.

The 1560.5 nm laser frequency is locked by injecting the 1560.5 nm infrared laser into a ultralow expansion (ULE) cavity (cavity length is 47.6 mm, free spectral range is 3.145 GHz, fineness is 34000@1560.5 nm, 30000@637.2 nm), and the laser frequency is locked by using the PDH sideband modulation technology. Here, a waveguide-type EOPM is added between the 1560.5 nm laser seed source and the amplifier for phase modulation of the laser, mainly because the EOPM cannot operate above the watt power due to its low damage threshold. On the other hand, the modulation frequency added to the 1560.5 nm laser can be transferred to the generated 637.2 nm red light through the sum-frequency process, which in turn enables the frequency-doubling cavity locking.

The frequency stabilization of 1076.9 nm laser is performed with the aid of 637.2 nm laser, using the electronic sideband (ESB) frequency stabilization technology$^{22}$, which is different from the PDH frequency stabilization by phase modulation of the modulation sideband carried by the 637.2 nm laser. Finally, the feedback signal is fed back to the PZT port of the 1076.9 nm fiber laser to realize the frequency stabilization of 1076.9 nm laser. This results in frequency stabilization of the entire 318.6 nm UV laser system, which has the advantage of continuous tuning (range >6 GHz) while locking the laser.

\begin{figure}
\includegraphics[width=8cm]{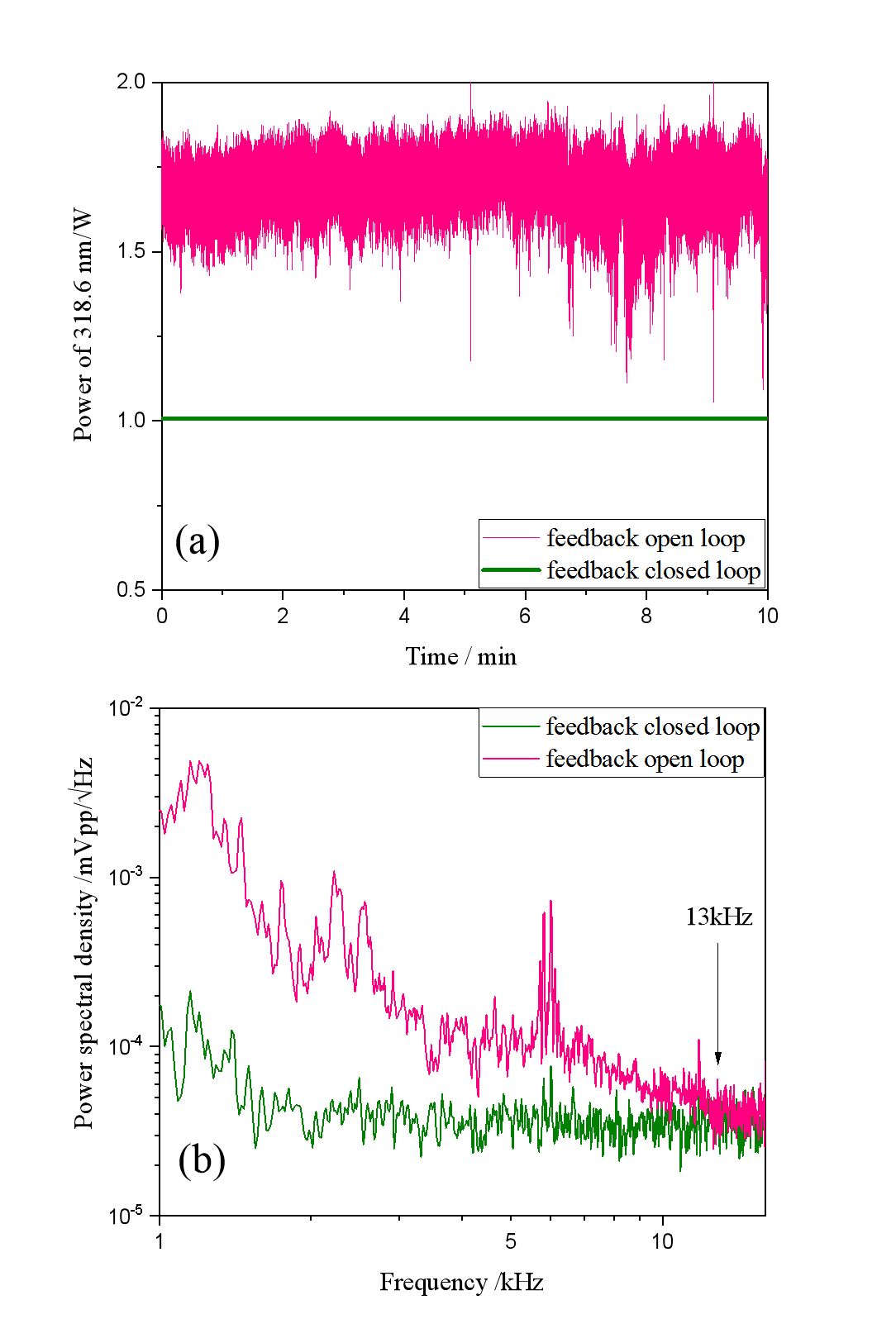}
\caption{\label{fig:epsart} When the feedback loop is open or closed, (a) the peak-peak value of power fluctuation in the time domain decreases from $\pm$ 12.50 $\%$ to $\pm$ 0.08 $\%$; (b) the effective bandwidth in the frequency domain is $\sim$ 13.0 kHz. }
\end{figure}

\begin{figure}
\includegraphics[width=8cm]{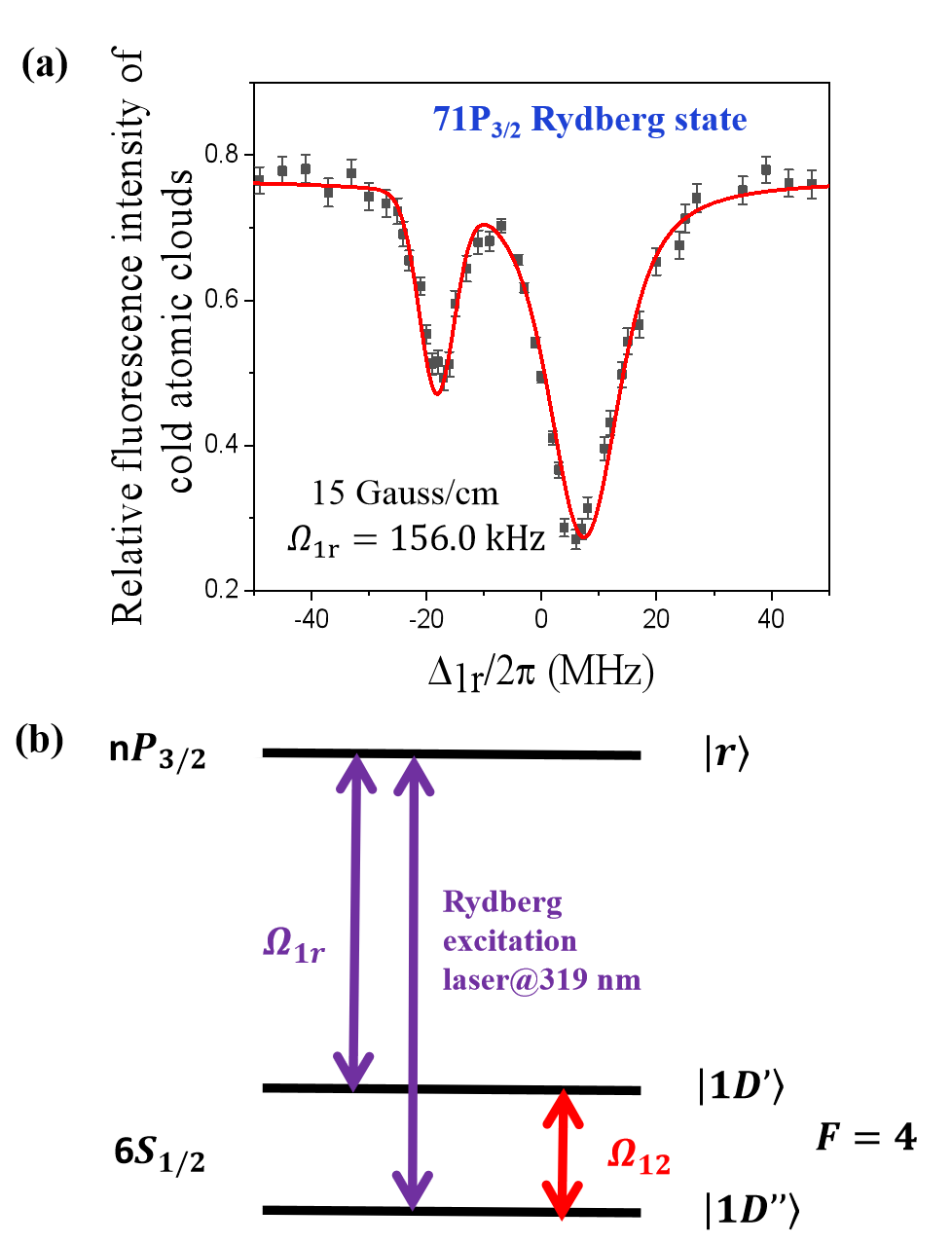}
\caption{\label{fig:epsart} (a) Trap-loss spectrum of  $6S_{1/2}(F=4)\rightarrow 71P_{3/2}$  Rydberg state, error bars are standard deviations obtained by means of multiple measurements; (b) Energy level of dressed states.When the strong coupling laser interacts with a two-level system, the atomic energy level is split into two which is induced by dressing splitting. Therefore, $6S_{1/2}(F=4)$ state forms two dressed-ground-states, $\mid {1D'}\rangle$ and $\mid {1D''}\rangle$, respectively. $\Omega_{1r}$ is the Rabi frequency of 318.6-nm UV laser, $\Delta_{12}$ is the detuning of 318.6-nm UV laser, excitation of Rydberg states is accomplished by a single-step direct excitation scheme from $6S_{1/2}(F=4)\rightarrow 71P_{3/2}$ state with a 318.6-nm UV laser. }
\end{figure}

The experimentally generated 318.6 nm laser is power stabilized by acousto-optic modulator(AOM), as shown in Fig.3, experimentally the 318.6 nm laser pass through the AOM and the 0-level light is blocked out and the -1-level diffracted light is taken, followed by a small angle sampling with BS, and the sampled laser pass through the servo control system after the detector and finally is used to control the AOM. The other laser beam after beam splitting is sampled by BS small-angle, one way for subsequent laser use, and one way through the detector divided into two parts for monitoring, using a digital multimeter to monitor the intensity fluctuations in the time domain, as well as the SR785 fast Fourier transform spectrometer to monitor the intensity fluctuations in the frequency domain. The laser intensity fluctuation in the time domain is suppressed from $\pm$ 12.50 $\%$ to $\pm$ 0.08 $\%$, the pink and green are the intensity fluctuation under the feedback open loop and closed loop, respectively, and the response bandwidth of the feedback loop in the frequency domain is about $\sim$ 13.0 kHz, as shown in Fig.4, after which the trap-loss spectrum is studied by combining the UV laser with a cesium magneto-optical trap. 

\section{Fluorescence trap-loss spectra of cesium cold atom samples in magneto-optical trap}

The single-step Rydberg excitation of cold cesium atoms is studied, experimentally, we adopt high-precision trap-loss spectroscopy to determine the Rydberg excitation, because Rydberg atoms cannot be trapped by MOT $^{23-24}$. In MOT, Rydberg excitation results in a reduction of the atom number in the 6S$_{1/2}$(F = 4) ground state, as atoms are excited to the Rydberg state. The cold atom fluorescence loss rate of atoms in MOT is proportional to the number of atoms excited to Rydberg state. The number of atoms excited to the Rydberg state can be estimated by measuring the cold atom fluorescence before and after the Rydberg excitation, to obtain the fluorescence loss rate. Specifically, we use a digital CCD camera (Thorlabs, 1500M-GE) to take spatially-resolved images of MOT and monitor the atom number of the ground state [6S$_{1/2}$(F = 4)]. MOT is continuously loaded, and the MOT fluorescence is recorded on the CCD with and without UV beam for 10 s each.

In the experiment, we use standing laser field excitation to reduce the radiation pressure of the laser, which can push cold atoms out of the magneto-optical trap. Since the UV laser is weak, the photoionization of cesium atoms due to UV laser should be relatively small. Therefore, the reduction of cold atoms in the magneto-optical trap is mainly due to the interaction of UV laser with cold atoms, which excites them from the ground state to the Rydberg state.

In the cold atom system, since MOT cannot capture Rydberg atoms, when 318.6 nm ultraviolet laser is applied to the cold cesium atoms in MOT, the atoms will be lost from the trap. Therefore, the excitation of Rydberg atoms is judged experimentally by trap-loss spectroscopy. That is, the fluorescence change of cold atom cloud caused by UV laser resonance with atoms is directly measured. The loss rate of cold atom fluorescence of atoms in MOT is proportional to the number of atoms excited to Rydberg state. The typical trap-loss spectrum is observed experimentally, corresponding to the $6S_{1/2}(F=4)\rightarrow 71P_{3/2}$ Rydberg state. As shown in Fig.5(a), $\Omega_{12}$ = 22.8 MHz, $\Omega_{1r}$ = 156.0 kHz, the double pits in the spectrum are due to AT splitting caused by cooling laser strong coupling. When the strong coupling light interacts with a two-level system (transition), the atomic energy level will dressed splitting, that is, the strong cooling laser interacts with the cold atom leads to the splitting of atomic energy levels, and the UV laser acting on the atomic level leads to new resonance. As shown in Fig. 5(b), the  $6S_{1/2}(F=4)$ state forms two dressed ground states($\mid {1D'}\rangle$ and $\mid {1D''}\rangle$), so that when the 318.6 nm UV laser acts on the atoms, two absorption pits are formed at the two dressed state positions.

\begin{figure*}
\includegraphics[width=16cm]{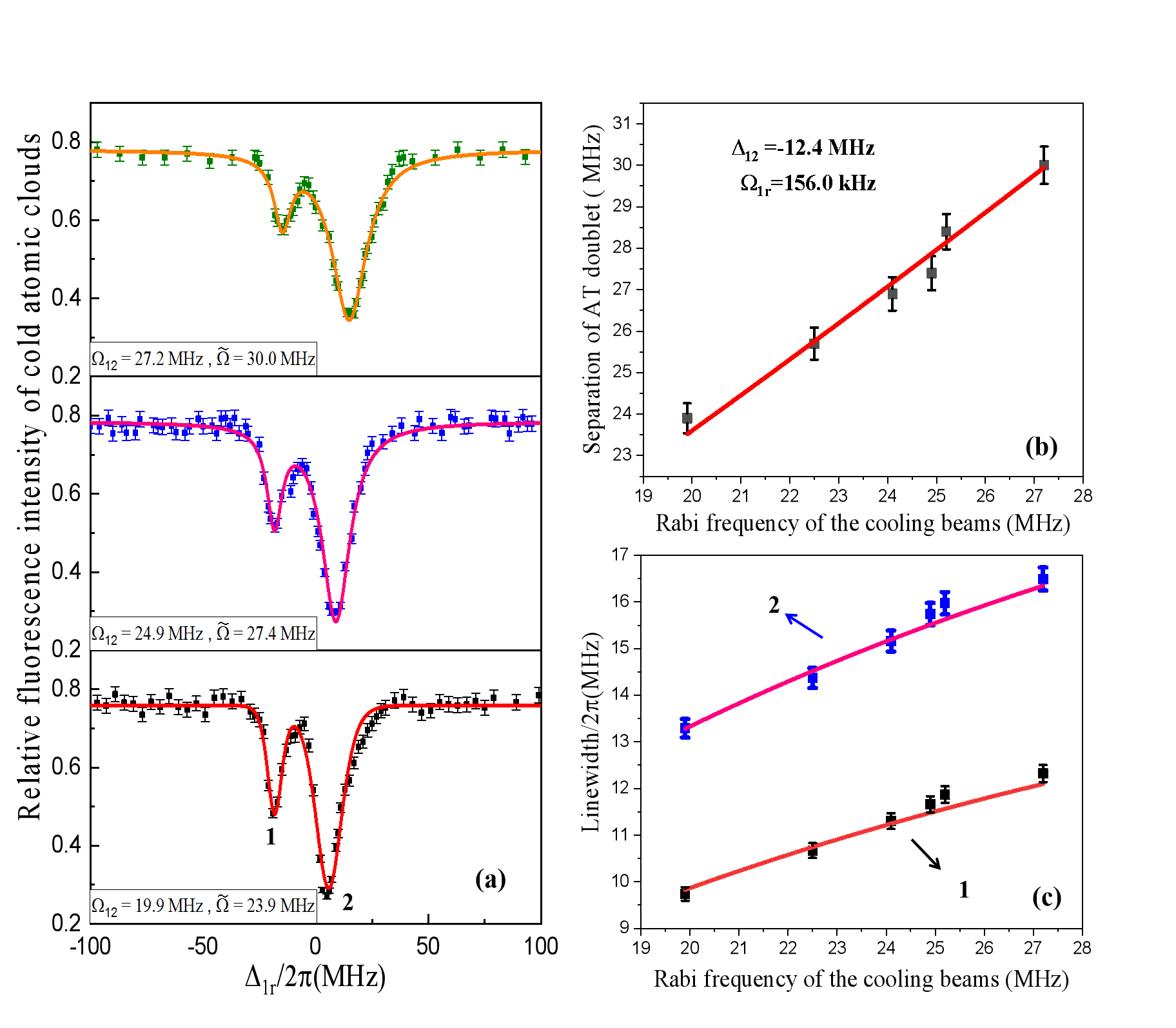}

\caption{\label{fig:wide}Trap-loss loss spectra are obtained by changing Rabi frequency of cooling laser; (a) The trap-loss spectra are observed when the Rabi frequencies of the cooling laser are 19.9, 24.9 and 27.2 MHz respectively. The solid lines are the results of theoretical calculations, and  AT splitting intervals are 23.9, 27.4 and 30.0 MHz respectively. (b) The AT split interval varies with the Rabi frequency of the cooling laser. (c) shows the linewidths of the AT splitting double pits as a function of the cooling laser Rabi frequency, respectively.}
\end{figure*}

Based on a frequency-stabilized tunable UV laser system, we measure the fluorescence trap-loss spectra of the 71P$_{3/2}$ Rydberg state at different Rabi frequencies of cooling laser. As shown in Fig. 6(a), here the UV power is fixed at $\sim$ 2 mW and the Rabi frequency of the cooling laser is varied, and the trap-loss spectrum is measured when the weak 318.6 nm UV laser is scanned near the $6S_{1/2}(F=4)\rightarrow 71P_{3/2}$ transition. Among them, Rabi frequencies of cooling laser corresponding to black, blue and green squares of data are 19.9, 24.9 and 27.2 MHz respectively, and the measured AT split intervals are 23.9, 27.4 and 30.0 MHz respectively, which are experimental measurement results. The solid line is based on the theoretical calculation of AT splitting spectra in a V-type three-level system. It can be seen from the figure that there are two pits in the spectrum, namely AT splitting. The AT splitting is caused by the strong coupling of the cooling laser. To further prove our inference, we also measure the variation of the AT interval with the Rabi frequency of cooling laser. As shown in Fig. 6(b), the black data are the experimentally measured extraction values, which are the AT splitting intervals extracted from the trap-loss spectra measured by changing Rabi frequencies of different cooling laser. It can be seen from the figure that the AT splitting intervals in the trap-loss spectra increase as the cooling light Rabi frequency increases, and the red solid line is the calculation result of the dressed state theory $^{25}$.

According to the theory of dressed states, the interval of the AT double pits can be expressed as $\tilde{\Omega}$ =$\sqrt{\Omega^2_{12}+\Delta^2_{12}}$, where $\Omega_{12}$ is the total Rabi frequency of the cooling laser and $\Delta_{12}$ = -12.4 MHz is the detuning of the cooling laser with respect to the $6S_{1/2}(F=4)\rightarrow 6P_{3/2}(F' = 5)$ hyperfine transition. The asymmetry of the AT double is due to the non-zero detuning of the cooling beam with respect to the hyperfine transition$^{25,26}$. We can see that the theoretical calculations are in general agreement with the experimental data, which further proves that the double pits are caused by the strong cooling laser leading to the dressed splitting of the cesium atom ground state.

\begin{figure*}
\includegraphics[width=16cm]{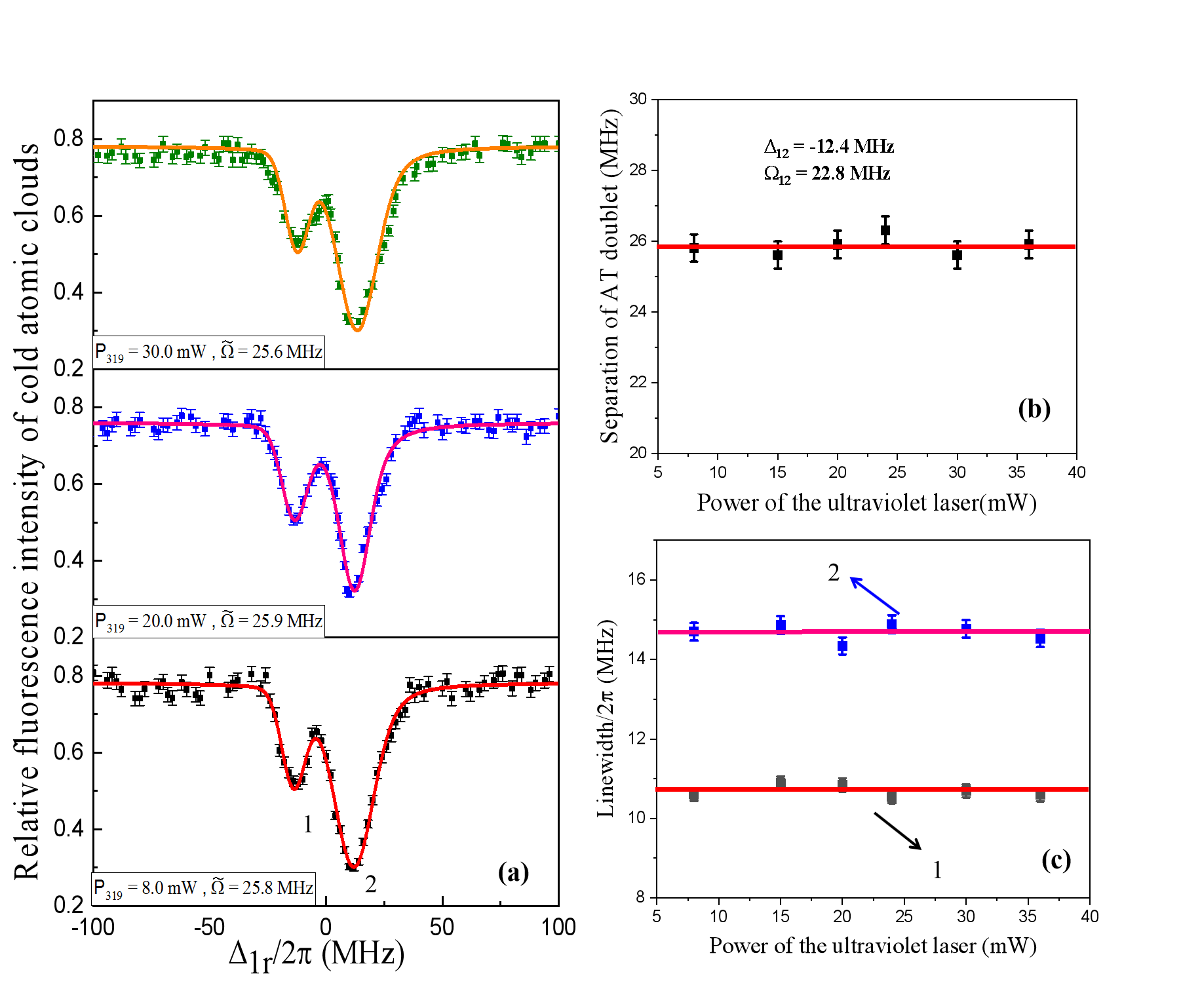}
\caption{\label{fig:wide}Trap-loss spectra are obtained by changing the power of the ultraviolet laser; (a) The trap- loss spectra are observed when the UV power are 8.0, 20.0 and 30.0 mW respectively. The AT splitting intervals are 25.8, 25.9 and 25.6 MHz respectively. (b) The AT split interval varies with the power of the ultraviolet laser.(c) shows the linewidths of AT split pits as a function of UV power.}
\end{figure*}

Fig.6(c) shows the linewidths of the AT splitting double pits as a function of the cooling laser Rabi frequency, respectively. According to reference$^{27}$, the linewidth of AT split double pits can be expressed as: $\Gamma_{\pm}$= $\frac{\Gamma+D}{2}$(1$\mp$$\frac{\Delta_{12}}{\sqrt{\Delta_{12}^2+4\Omega^2_{12}}}$)+W, where  $\Gamma_{\pm}$ is linewidth of two states, D is the Doppler broadening, W is the other broadening mechanism, and $\Gamma$ is the decay rate from the excited state to the ground state. The spectral linewidth is mainly affected by Doppler broadening, power broadening, interatomic collision broadening, and transition broadening, etc. The squares are the experimental data and the solid lines are the fitting results.

The fluorescence trap-loss spectra of the $6S_{1/2}(F=4)\rightarrow 71P_{3/2}$ Rydberg state under varying UV power conditions are measured in Fig. 7(a). The black, blue, and green squares of the data correspond to the UV power of 8.0, 20.0, and 30.0 mW, respectively, and the measured AT splitting intervals are 25.8, 25.9, and 25.6 MHz, which are the experimental measurement results. Here, the Rabi frequency $\Omega_{12}$ of the cooling laser is 22.8 MHz, and the frequency detuning $\Delta_{12}$ is -12.4 MHz. It can be seen from the figure that the AT splitting interval changes little with the change of UV power. This is because the Rabi frequency of the UV laser is much smaller than the Rabi frequency of the cooling laser, so the AT splitting interval basically does not vary with the power of weak UV laser, as shown in Fig. 7(b), we measure the AT splitting interval for six groups of UV laser power varying from 5-40 mW, respectively, and it can be seen from the figure that the AT splitting interval is about 25.8(2) MHz, which is basically consistent with the theoretical calculation value of 25.9 MHz. Fig. 7(c) shows the linewidths of AT split pits as a function of UV power, from the figure, it can be seen that the linewidth basically does not vary with the UV power.

\section{Conclusion}

In the paper, a single-step Rydberg excitation experiment of the cesium cold atomic system is carried out by a power-stabilized 318.6 nm laser system. Based on the trap-loss spectroscopy technology, we realize the nondestructive detection of the Rydberg state and observe the Autler-Townes splitting in the cold atom ensemble due to the strong cooling laser, and investigate the parameter dependence of the AT splitting interval in the trap-loss spectroscopy, which satisfy the theory of dressed states. Furthermore, the relevant parameters of cold atom samples are measured in the cesium magneto-optical trap, including the size, effective temperature, and atomic number density of the cold atoms. It has positive significance for further development of quantum information processing and quantum computing by using single-step Rydberg excitation in cold atom system.

\section*{Funding}

This research is partially funded by the National Key R$\&$D Program of China (2021YFA1402002), the National Natural Science Foundation of China ( 61875111 and 12104417),  and the Fundamental Research Program of Shanxi Province (20210302124161).

\section*{Conflict of Interest}

The authors have no conflicts to disclose.

\nocite{*}
\bibliography{aipsamp}

\section*{References} 

$^1$ A. Browaeys, T. Lahaye. Many-body physics with individually controlled Rydberg atoms[J]. Nature Phys., {\bf 16}, 132-142, (2020).\\
$^2$ C. S. Adams, J. D. Pritchard, J. P. Shaffer, Rydberg atom quantum technologies[J], J. Phys. B: At. Mol. Opt. Phys.,{\bf 53}, 012002, (2020).\\
$^3$ M. Saffman, T. G. Walker, K. Mølmer. Quantum information with Rydberg atoms[J]. Rev. Mod.  Phys., {\bf 82}, 2313-2363, (2010).\\
$^4$ A. V. Zasedatelev, A. V. Baranikov, D. Sannikov, D. Urbonas, F. Scafirimuto, V. Yu. Shishkov, E. S. Andrianov, Y. E. Lozovik, U. Scherf, T. Stöferle, R. F. Mahrt, and P. G. Lagoudakis, Single-photon nonlinearity at room temperature[J], Nature, {\bf 597}, 493-497, (2021).\\
$^5$ M. Moreno-Cardoner, D. Goncalves, D. E. Chang, Quantum nonlinear optics based on two-dimensional Rydberg atom arrays[J]. Phys. Rev. Lett., {\bf 127}, 263602, (2021).\\
$^6$ M. Ferreira-Cao, V. Gavryusev, T. Franz, R. F. Alves, A. Signoles, G. Zürn and M. Weidemüller, Depletion imaging of Rydberg atoms in cold atomic gases[J]. J. Phys. B: At. Mol. Opt. Phys., {\bf 53}, 084004, (2020).\\
$^7$ C. Gross, T. Vogt, W. Li, Ion imaging via long-range interaction with Rydberg atoms[J]. Phys. Rev. Lett., {\bf 124}, 053401, (2020).\\
$^8$ B. S. Silpa, A. B. Shovan, C. Saptarishi, and R. Sanjukta, Transition frequency measurement of highly excited Rydberg states of $^{87}$Rb for a wide range of principal quantum numbers[J]. Optics Continuum, {\bf 1}, 1176-1192, (2022).\\
$^9$ J. M. Zhao, H. Zhang, Z. G. Feng, X. B. Zhu, L. J. Zhang, C. Y. Li, and S. T. Jia, Measurement of polarizability of cesium nD state in magneto-optical trap[J]. J. Phys. Soc. Japan, {\bf 80}, 034303, (2011).\\
$^{10}$ W. C. Xu, A. V. Venkatramani, S. H. Cantú, T. Šumarac, V. Klüsener, M. D. Lukin, and V. Vuletić, Fast preparation and detection of a Rydberg qubit using atomic ensembles[J], Phys. Rev. Lett., {\bf 127}, 050501, (2021).\\
$^{11}$ T. Firdoshi, S. Garain, V. Gupta, D. Kara, and A. K. Mohapatra, Electromagnetically induced transparency in the strong blockade regime using the four-photon excitation process in thermal rubidium vapor[J], Phys. Rev. A, {\bf 104}, 013711, (2021).\\
$^{12}$ D. Tong, S. M. Farooqi, J. Stanojevic, S. Krishnan, Y. P. Zhang, R. Côté, E. E. Eyler, and P. L. Gould, Local blockade of Rydberg excitation in an ultracold gas[J]. Phys. Rev. Lett., {\bf 93}, 063001, (2004). \\
$^{13}$ P. Thoumany, T. Hänsch, G. Stania, L. Urbonas, and Th. Becker, Optical spectroscopy of rubidium Rydberg atoms with a 297 nm frequency-doubled dye laser[J]. Opt. Lett., {\bf 34}, 1621-1623, (2009).\\
$^{14}$ A. Arias, G. Lochead, T. M. Wintermantel, S. Helmrich, and S. Whitlock, Realization of a Rydberg-dressed ramsey interferometer and electrometer[J], Phys. Rev. Lett., {\bf 122} 053601, (2019).\\
$^{15}$ A. M. Hankin, Y. Y. Jau, L. P. Parazzoli, C. W. Chou, D. J. Armstrong, A. J. Landahl, and G. W. Biedermann, Two-atom Rydberg blockade using direct 6S to nP excitation[J]. Phys. Rev. A, {\bf 89}, 033416, (2014).\\
$^{16}$ M. Li, B. Li, X. Jiang, J. Qian, X. Li, and L. Liu, Measurement of $^{85}$Rb nP-state transition frequencies via single-photon Rydberg excitation spectroscopy[J]. J. Opt. Soc. Am. B, {\bf 36}, 1850-1857, (2019).\\
$^{17}$ B. Li, M. Li, X. Jiang, J. Qian, X. Li, L. Liu, and Y. Wang, Optical spectroscopy of nP Rydberg states of $^{87}$Rb atoms with a 297-nm ultraviolet laser[J]. Phys. Rev. A, {\bf 99}, 042502, (2019).\\
$^{18}$ J. D. Bai, S. Liu, J. Y. Wang, J. He, and J. M. Wang, Single-photon Rydberg excitation and trap-loss spectroscopy of cold cesium atoms in a magneto-optical trap by using of a 319-nm ultraviolet laser system[J], IEEE J. Sel. Top. Quant. Electr., {\bf 26}, 1600106, (2020).\\
$^{19}$ J. Y. Wang, J. D. Bai, J. He, and J. M. Wang, Realization and characterization of single-frequency tunable 637.2nm high-power laser[J], Opt. Commun., {\bf 370}, 150-155, (2016).\\
$^{20}$ J. M. Wang, J. D. Bai, J. Y, Wang, and J. He, Realization of a watt-level 319-nm single-frequency CW ultraviolet laser and its application in single-photon Rydberg excitation of cesium atoms[J], Chinese Optics, {\bf 12}, 701-718, (2019). (in Chinese)\\
$^{21}$ J. Y. Wang, J. D. Bai, J. He, and J. M. Wang, Development and characterization of a 2.2 W narrow-linewidth 318.6 nm ultraviolet laser[J], J. Opt. Soc. Am. B, {\bf 33}, 2020-2025, (2016).\\
$^{22}$ J. D. Bai, J. Y. Wang, J. He, and J. M. Wang, Electronic sideband locking of a broadly tunable 318.6 nm ultraviolet laser to an ultra-stable optical cavity[J]. J. Opt., {\bf 19}, 045501, (2017). \\
$^{23}$ A. G. Boetes, R. V. Skannrup, J. B. Naber, S. J. J. M. F. Kokkelmans, and R. J. C. Spreeuw, Trapping of Rydberg atoms in tight magnetic microtraps[J]. Phys. Rev. A, {\bf 97}, 013430, (2018).\\
$^{24}$ Y. F. Cao, W. G. Yang, H. Zhang, M. Y. Jing, W. B. Li, L. J. Zhang, L. T. Xiao, and S. T. Jia, Dephasing effect of Rydberg states on trap loss spectroscopy of cold atoms[J], J. Opt. Soc. Am. B, {\bf 39}, 2032-2036, (2022).\\
$^{25}$ S. Menon, and G. S. Agarwal. Gain components in the Autler-Townes doublet from quantum interferences in decay channels[J]. Phys. Rev. A, {\bf 61}, 013807, (1999).\\
$^{26}$ J. D. Bai, J. Y. Wang, S. Liu, J. He, and J. M. Wang. Autler-Townes doublet in single-photon Rydberg spectra of cesium atomic vapor with a 319 nm UV laser[J]. Appl. Phys. B, {\bf 125}, 33, (2019).\\
$^{27}$ G. S. Agarwal. Nature of the quantum interference in electromagnetic-field-induced control of absorption. Phys. Rev. A, {\bf 55}, 2467, (1997).\\

\end{document}